\documentclass{article}

\usepackage{PRIMEarxiv}

\usepackage[utf8]{inputenc} % allow utf-8 input
\usepackage[T1]{fontenc}    % use 8-bit T1 fonts
\usepackage{hyperref}       % hyperlinks
\usepackage{url}            % simple URL typesetting
\usepackage{booktabs}       % professional-quality tables
\usepackage{amsfonts}       % blackboard math symbols
\usepackage{nicefrac}       % compact symbols for 1/2, etc.
\usepackage{microtype}      % microtypography
\usepackage{lipsum}
\usepackage{fancyhdr}       % header
\usepackage{graphicx}       % graphics
\graphicspath{{media/}}     % organize your images and other figures under media/ folder

\usepackage{setspace}   % Allows to control linespacing
\usepackage{multirow} % Allows to merge table cells at will
\usepackage[numbers]{natbib}

\usepackage{caption}
\usepackage{subcaption}
\usepackage{xcolor}

\newcommand{\revision}[1]{\textcolor{black}{#1}}

\title{Interactive visualization of large molecular systems with VTX: example with a minimal whole-cell model
\thanks{\textit{\underline{Citation}}: 
\textbf{Maxime Maria, Valentin Guillaume, Simon Guionnière, Nicolas Dacquay, Cyprien Plateau–Holleville,  Vincent Larroque, Jean Lardé, Yassine Naimi, Jean-Philip Piquemal, Guillaume Levieux, Nathalie Lagarde, Stéphane Mérillou and Matthieu Montes. Interactive visualization of large molecular systems with VTX: example with a minimal whole-cell model, DOI : 10.3389/fbinf.2025.1588661}} 
}

\begin{document}
\maketitle

Maxime Maria \textsuperscript{1}, Valentin Guillaume \textsuperscript{2}, Simon Guionnière \textsuperscript{2}, Nicolas Dacquay \textsuperscript{2}, Cyprien Plateau–Holleville \textsuperscript{1},  Vincent Larroque \textsuperscript{1,3}, Jean Lardé \textsuperscript{2,3}, Yassine Naimi \textsuperscript{3}, Jean-Philip Piquemal  \textsuperscript{4,5,6}, Guillaume Levieux \textsuperscript{7}, Nathalie Lagarde \textsuperscript{2}, Stéphane Mérillou \textsuperscript{1} and Matthieu Montes \textsuperscript{2,6,*}

\begin{spacing}{0.75}
\begingroup
    \tiny 
    \textsuperscript{1}XLIM, UMR CNRS 7252, Université de Limoges, 87000, Limoges, France, 
    \textsuperscript{2}Laboratoire GBCM, EA 7528, Conservatoire National des Artset Métiers, 75003, Paris, France, 
    \textsuperscript{3}Qubit Pharmaceuticals SAS, France, 
    \textsuperscript{4}LCT, UMR 7616 CNRS, Sorbonne Université, Paris, France,
    \textsuperscript{5}Department of Biomedical Engineering, University of Texas at Austin, Texas, USA, 
    \textsuperscript{6}Institut Universitaire de France, Paris, France and
    \textsuperscript{7}Laboratoire CEDRIC, EA 4626, Conservatoire National des Arts et Métiers, 75003, Paris, France.\\
    *Corresponding authors: maxime.maria@unilim.fr, matthieu.montes@cnam.fr\\
\endgroup
\end{spacing}

\begin{abstract}
    VTX is an open-source molecular visualization software designed to overcome the scaling limitations of existing real-time molecular visualization software when handling massive molecular datasets. VTX employs a meshless molecular graphics engine utilizing impostor-based techniques and adaptive level-of-detail (LOD) rendering. This approach significantly reduces memory usage and enables real-time visualization and manipulation of large molecular systems. Performance benchmarks against VMD, PyMOL, and ChimeraX using a 114-million-bead Martini minimal whole-cell model demonstrate VTX's efficiency, maintaining consistent frame rates even under interactive manipulation on standard computer hardware. VTX incorporates features such as screen-space ambient occlusion (SSAO) for enhanced depth perception and free-fly navigation for intuitive exploration of large molecular systems. VTX is open-source and free for non commercial use. Binaries for Windows and Ubuntu Linux are available at \href{http://vtx.drugdesign.fr}{http://vtx.drugdesign.fr}. VTX source code is available at \href{https://github.com/VTX-Molecular-Visualization}{https://github.com/VTX-Molecular-Visualization}.

\end{abstract}


\section*{Introduction}

Recent advances in the determination of atomic resolution molecular structures and assemblies using CryoEM~\cite{cryoem}, in protein structure prediction methods like AlphaFold~\cite{alphafold,alphafolddb}, and the growing availability of molecular dynamics simulations~\cite{amaro24,hospital24}, have significantly increased the volume of molecular structural biology data. 
Additionally, modern high-performance computing hardware has enabled the simulation of larger molecular systems\revision{~\cite{sanbonmatsu,marrink23,aksimentiev24, Gilbert2023}}, further amplifying data production. This massive molecular data poses a challenge for existing software solutions in terms of data processing, visualization, and storage. 

In this work, we demonstrate the performance of VTX~\cite{VTX} on large molecular datasets by visualizing a 114 million Martini-beads minimal whole cell model published in 2023~\cite{marrink23}, which pre-figures future massive molecular systems for molecular dynamics simulation. We describe the solutions implemented in VTX to efficiently handle large all-atom molecular data while ensuring high-quality and informative real-time rendering. This serves as a valuable benchmark for future applications of massive molecular dynamics simulation visualization, where memory management and high-quality rendering remain key challenges.

\section*{Methods}\label{sec:method}

VTX~\cite{VTX} is an open-source molecular visualization software designed for efficient real-time handling of large molecular simulations datasets. 
It achieves this through a notably high-performance, meshless molecular graphics engine coupled with a minimalistic, task-oriented GUI. Here, we illustrate VTX's features and performance on a massive molecular system: the 2023 Martini minimal whole cell model~\cite{marrink23}.  

\subsection*{Molecular Graphics Engine}

\textit{\textbf{Meshless representations.}} 
All molecular representation can be described using a triangular mesh. To draw a triangular mesh, each triangle requires the XYZ coordinates of its three points, resulting in at least 36 bytes of memory per triangle. 

\begin{figure*}[hbt!]
    \centering 
    \begin{subfigure}[b]{0.79\textwidth}
         \centering
         \includegraphics[width=\textwidth]{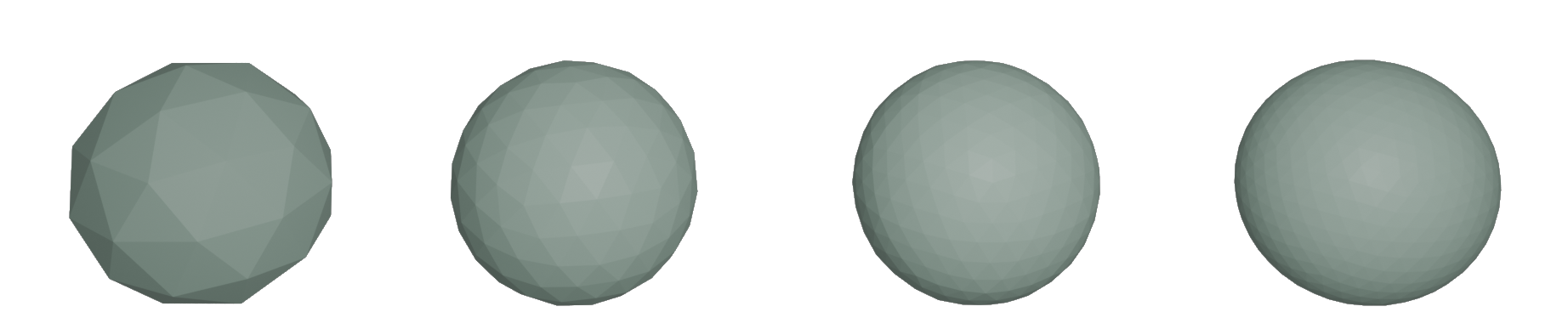}
         \caption{Triangle mesh-based spheres\label{fig:mesh_a}}
     \end{subfigure}
    \begin{subfigure}[b]{0.20\textwidth}
         \centering
         \includegraphics[width=\textwidth]{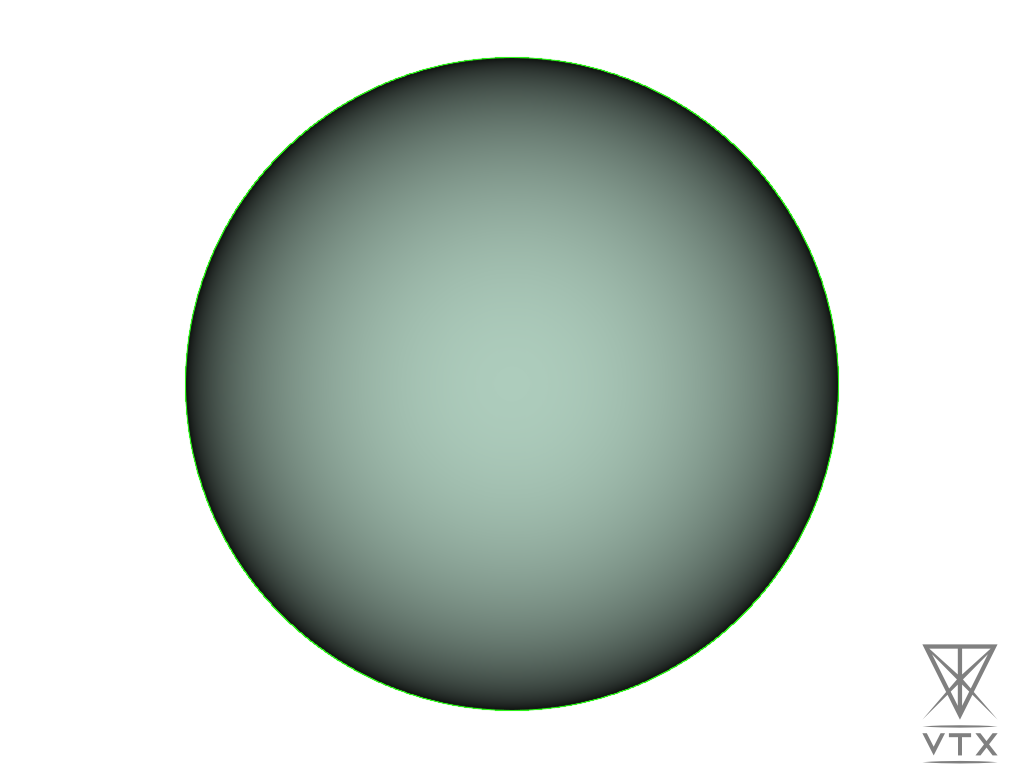}
         \caption{Impostor-based sphere\label{fig:mesh_b}}
     \end{subfigure}
    \caption{Illustration of the complexity of mesh-based and impostor-based sphere representations. (a)~Mesh-based spheres with increasing triangle counts (80, 320, 720, and 1280 respectively) generated with Blender; (b)~VTX's impostor-based sphere (with only one vertex).}

    \label{fig:mesh}
\end{figure*}

In order to produce images of acceptable quality, many triangles are needed to accurately define the geometry and have smooth surfaces (see figure~\ref{fig:mesh_a}).
Even if GPUs are designed for fast triangle rendering, the number of required triangles is prohibitive for real-time visualization.
 
To address this issue, \textit{Ball and Stick} and \textit{Van der Waals} representations are based on quadrics such as spheres or cylinders that
are described implicitly, using atom coordinates and radii exclusively.
To achieve this, the rendering engine employs impostor-based techniques~\cite{impostors}. 
For each primitive, a simple quad is rasterized. 
Then, ray-casting is used to evaluate the implicit equation of the primitive, generating the final shape. 
This approach enables fast and pixel-perfect rendering (see figure~\ref{fig:mesh_b}) while minimizing both bandwidth and memory usage, which is crucial for handling large molecular structures or dynamics data.

\textit{\textbf{Cartoon Representation.}} 
Cartoon representations in VTX use an adaptive level-of-detail (LOD) approach with tessellation shaders~\cite{cartoonSS}. Triangles are dynamically computed for each frame based on the atom coordinates, enabling fast rendering of secondary structure cartoons without requiring any preprocess. This allows updates at each trajectory step with minimal computational cost.

\textit{\textbf{Solvent Excluded Surface (SES).}} As is common in molecular visualization software, the surface is simply computed using the marching cubes algorithm~\cite{marchingcubes}.
This approach requires balancing surface smoothness and the number of generated triangles to ensure both visual quality and computational efficiency.

\textit{\textbf{No instancing.}} Instancing is a widely used technique for rendering multiple copies of the same object in a scene simultaneously, significantly improving performance when dealing with many identical objects. By reducing the number of draw calls, it saves both CPU and GPU resource usage. Massive static molecular scenes, containing up to billion atoms, as displayed with YASARA~\cite{yasaraview} or Mol*~\cite{molstar, mesoscale} effectively leverage instancing. However, instancing assumes that all copies share the same conformation, making it unsuitable for molecular dynamics trajectories where structures evolve independently. In this work, which pre-figures future applications for visualizing molecular dynamics trajectories of massive molecular systems, no instancing is used to display the 2023 Martini minimal whole cell model~\cite{marrink23}.  

\begin{figure*}[hbt!]
    \centering 
    \begin{subfigure}[b]{0.32\textwidth}
         \centering
         \includegraphics[width=\textwidth]{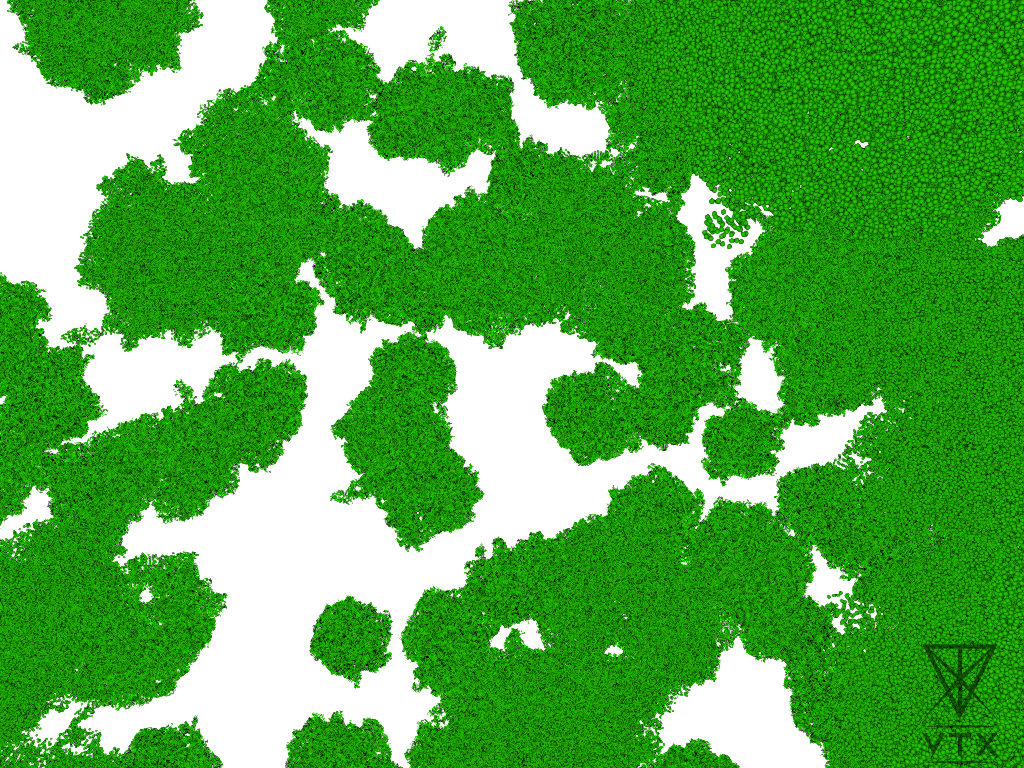}
         \caption{No SSAO}
         \label{fig:wo-SSAO}
     \end{subfigure}
    \begin{subfigure}[b]{0.32\textwidth}
         \centering
         \includegraphics[width=\textwidth]{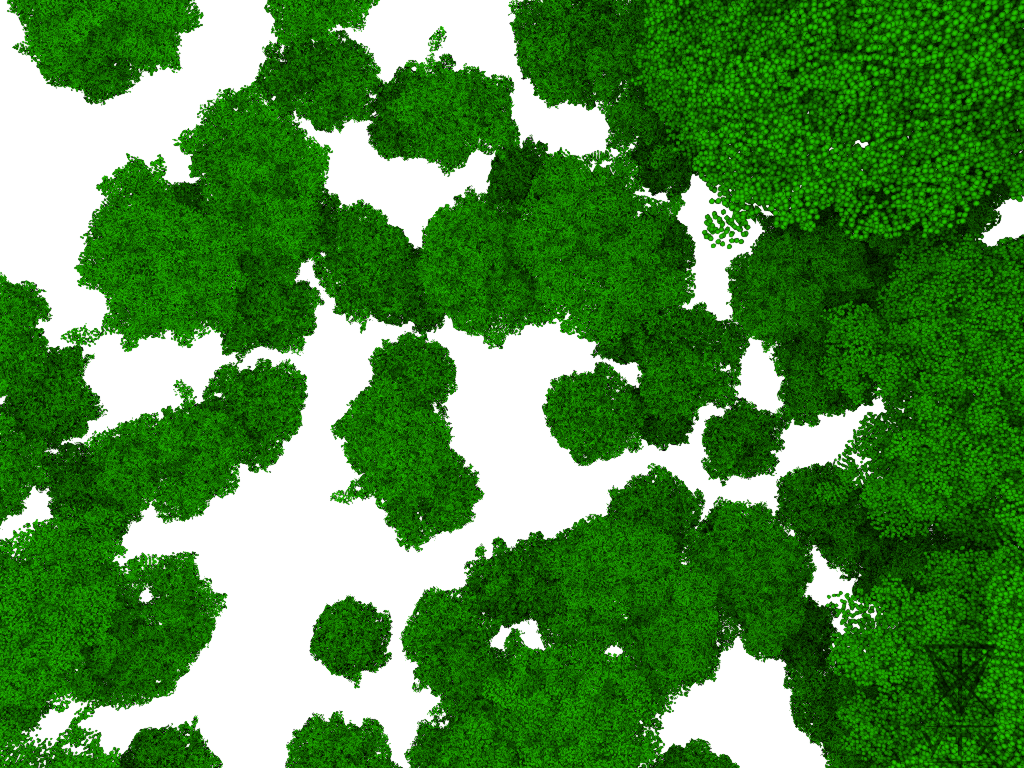}
         \caption{Low intensity SSAO}
         \label{fig:lil-SSAO}
     \end{subfigure}
    \begin{subfigure}[b]{0.32\textwidth}
         \centering
         \includegraphics[width=\textwidth]{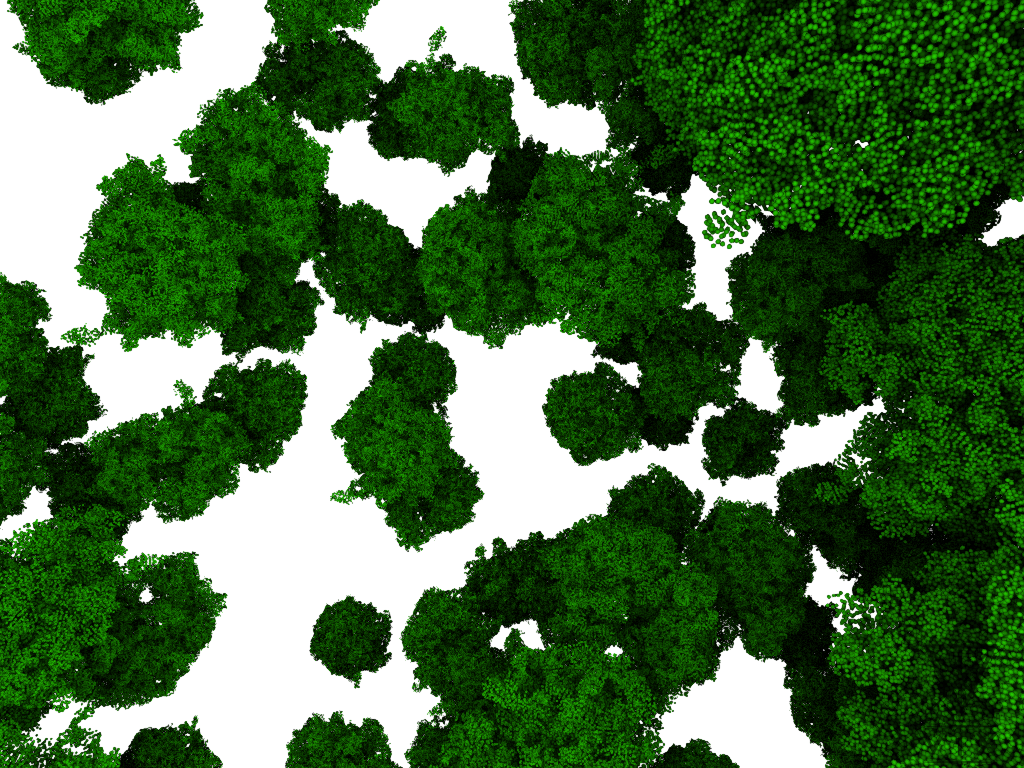}
         \caption{High Intensity SSAO}
         \label{fig:normal-SSAO}
     \end{subfigure}
    
    \caption{Impact of the intensity of the Screen Space Ambient Occlusion (SSAO) on a close-up rendering of a large molecular system~\cite{marrink23}~: (a)~No SSAO; (b)~Low intensity SSAO; (c)~High intensity SSAO.}
    \label{fig:SSAO}
\end{figure*}

\textit{\textbf{Lighting.}} Depth perception is essential for localizing and distinguishing molecular entities in a complex scene.  For this purpose, VTX uses a Screen-Space Ambient Occlusion (SSAO)~\cite{mittring07} to enhance the perception of structural details and emphasize the burriedness of atoms, improving the understanding of molecular architectures (see figure~\ref{fig:SSAO}).

\subsection*{Free-fly Navigation and interaction}

Navigating through large systems composed by multiple proteins, lipids and other heavy molecules poses a new challenge compared to single-protein scale visualization. 
Traditional trackball navigation, which rotates the camera around a fixed point, can be restrictive for large systems, making it difficult to explore surrounding regions.

To address this issue, VTX implements a free-fly camera mode in addition to the default trackball. Free-fly mode allows to move freely within the 3D space using keyboard inputs for movement 
and the mouse controls for directions, similar to the first-person navigation in video games. This approach has successfully applied in interactive docking tool UDock~\cite{levieux2014, udock2}, and in the multi-scale molecular visualization software CellVIEW~\cite{cellview}.\\

\subsection*{Performance evaluation on a massive molecular system: the 2023 Martini minimal whole cell model}

We compared the performance of VTX version 0.4.4~\cite{VTX}, VMD version 1.9.3~\cite{vmd}, PyMOL version 3.1~\cite{pymol} and ChimeraX version 1.9~\cite{chimeraX} in terms of system file loading time and frame rate using the 2023 Martini minimal whole cell model \cite{marrink23}. This system is a coarse-grained model of 114 million Martini beads available in the Gromacs~\cite{gromacs} \textit{.gro} file format that pre-figures future massive molecular systems constructed for molecular dynamics simulation. 
The system, displayed in figure~\ref{fig:rendering} contains 60,887 soluble proteins, 2,200 membrane proteins, 503 ribosomes, a single 500 kbp circular dsDNA, 1.3 million lipids, 1.7 million metabolites and 14 million ions. For clarity of the visual analysis, we did not display the 447 million water beads. 

The benchmark results are presented in table~\ref{table:bench}.
ChimeraX and PyMOL were unable to load the system due respectively to a crash and a freeze.
VMD and VTX loaded the system with similar times and successfully displayed it, but VTX achieved significantly higher frame rate.
Any attempts to modify VMD rendering settings caused in a severe drop in frame rate, eventually leading to a complete freeze.
With a stable interactive frame rate and despite limited computational resource, VTX allowed to interactively manipulate the system, perform precise selections and modify rendering settings.

A video displaying a free-fly real-time navigation inside this system is available as a supplementary material.

\begin{table}[hbt!]
    \centering 
    \begin{tabular}{c|c|c|c|}
    \cline{2-4}
     & Loading time (s) & Frame rate - close-up (s\textsuperscript{-1}) & Frame rate - far (s\textsuperscript{-1})\\ \hline
    \multicolumn{1}{|c|}{ChimeraX} & / & / & / \\ \hline
     \multicolumn{1}{|c|}{PyMOL3.1} & / & / & / \\ \hline
    \multicolumn{1}{|c|}{ VMD} & 200.33 ± 16.05 & 1.36 & 1.38\\ \hline
     \multicolumn{1}{|c|}{VTX} & 205.00 ± 13.06 & 11.41 & 12.82\\ \hline
    \end{tabular}
    \caption{Loading time and frame rate (benchmark results on the 2023 Martini minimal whole cell model (114 million Martini beads)~\cite{marrink23}. No values could be obtained with ChimeraX and PyMOL due to a frozen application (ChimeraX) or a crash (PyMOL) during system loading. System file loading time performance has been evaluated in triplicate. Frame rate measurement has been evaluated using the NVidia Frameview software version 1.6. The average frame rate was calculated over a 20-second period for both close-up (as in figure 2) and far (as in figure 3) camera views. All performance evaluations were conducted on a Dell precision 5480 laptop equipped with an Intel i7-13800H, 32GB of RAM and a NVidia Quadro RTX 3000 GPU. 
}
\label{table:bench}
\end{table}

\begin{figure*}[hbt!]
    \centering 
    \begin{subfigure}[b]{0.45\textwidth}
         \centering
         \includegraphics[width=\textwidth]{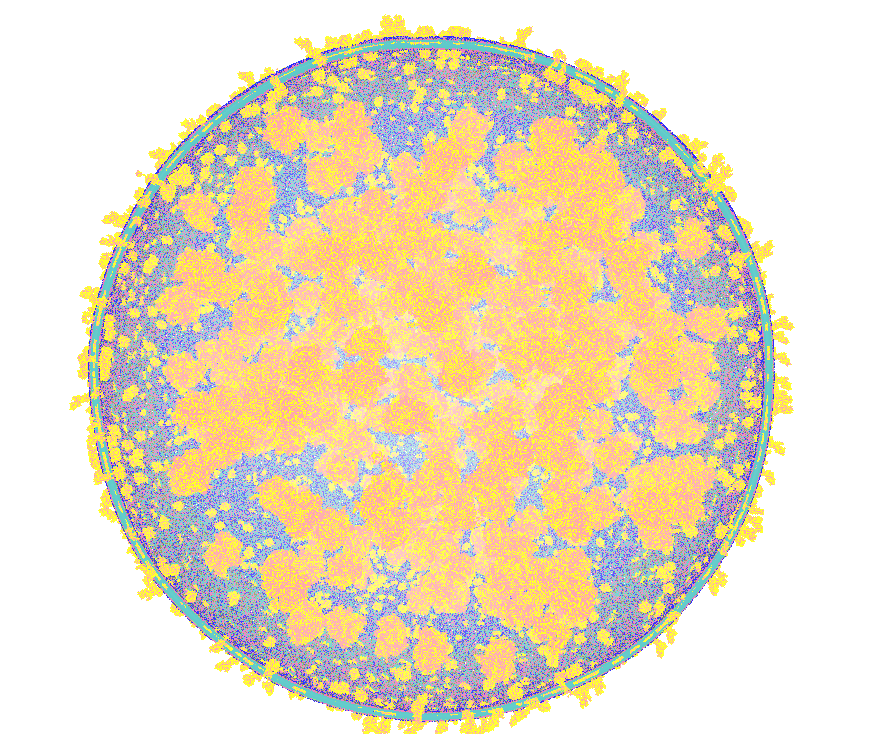}
         \caption{VMD scene}
         \label{fig:vmd_opening_view}
     \end{subfigure}
    \begin{subfigure}[b]{0.50\textwidth}
         \centering
         \includegraphics[width=\textwidth]{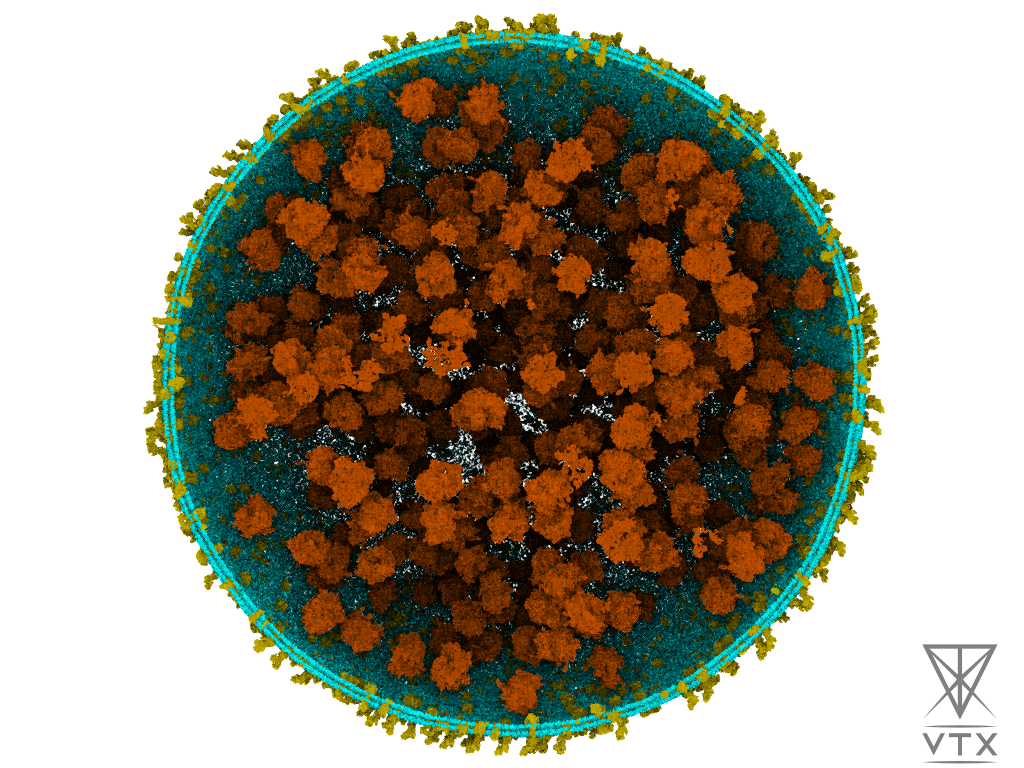}
         \caption{VTX scene}
         \label{fig:vtx_clip}
     \end{subfigure}
    \caption{Wide-system view of \revision{the lipids, membrane proteins, and ribosomes (36,566,468 Martini beads) of the 2023 Martini minimal whole cell model rendered in real-time with VMD on the left, and VTX on the right. These illustrations were captured on a Dell precision 5480 laptop equipped with an Intel i7-13800H, 32GB of RAM and a NVidia Quadro RTX 3000 GPU. All attempts to improve the rendering with VMD on this setup resulted in a freeze of the application}}
\end{figure*}

\section*{Perspectives}

\revision{VMD, ChimeraX and Py\MakeUppercase{MOL} are reference molecular visualization tools that are widely used by the community and that, compared to VTX,  offer many additional functionalities notably through the use of numerous and diverse plugins developed over time. In the next version of VTX, Python bindings will be supported to allow the development and integration of external scripts and plugins.} 

The construction and rendering of a Solvent Excluded Surface (SES) for large molecular systems currently exceed the capabilities of existing molecular visualization software due to high memory demands. To adress this limitation, we will integrate our analytical SES construction method~\cite{SEScyprien} into VTX. This approach will also enable the display of a smooth, pixel-perfect molecular surface.

Visualizing large system trajectories presents a challenge due to the memory demands of loading all atomic coordinates for each frame. Currently, VTX loads all frames to ensure smooth playback, which can cause performance issues on memory-constrained systems. To address this, we will implement a streaming system that limits the number of frames loaded into memory simultaneously, significantly reducing the memory footprint for large systems.

Even if VTX can already render large molecular systems without relying on instancing, further performance gains could be achieved by enabling instancing for identical entities. This would leverage VTX's graphics engine to display systems containing billions of atoms. For certain applications, such as visualizing repetitive molecular assemblies or large-scale structural models, instancing could significantly enhance rendering efficiency and reduce memory usage.

Finally, we are developing a ray tracing engine specifically designed for molecular scenes, integrating state-of-the-art computer graphics techniques to enable the fast generation of high-quality images and interactive previews that accurately reflect the final rendering.

\section*{Conclusion} 

VTX is an open-source, high performance molecular visualization software designed for real-time exploration of massive molecular systems. Its graphics engine, leveraging impostor-based rendering and adaptive level-of-detail, achieves interactive frame rates with a 114-million-atom model, outperforming existing tools on standard hardware. 
VTX integrates features such as screen-space ambient occlusion for enhanced depth perception and free-fly navigation for intuitive exploration of large systems. 
Future developments will include analytical Solvent-Excluded Surface construction, a streaming system for handling large molecular dynamics trajectories, instancing for reducing the memory load with multiple rigid copies of the same molecular object, and high-quality path-traced rendering. 
As an open-source software, VTX nature fosters community-driven advancements in large-scale molecular structure and dynamics visualization. 

\section*{Data and software availability}

VTX is open-source and free for non commercial use under the VTX consortium license. Builds for Windows and Linux are available at \href{http://vtx.drugdesign.fr}{http://vtx.drugdesign.fr}. The source code is available at \href{https://github.com/VTX-Molecular-Visualization}{https://github.com/VTX-Molecular-Visualization}.

\revision{All .gro files that constitute the Martini minimal whole cell model are available upon request at the Marrink lab. The procedure to generate the model is available at \href{https://github.com/marrink-lab/Martini\_Minimal\_Cell}{https://github.com/marrink-lab/Martini\_Minimal\_Cell}.}

\section*{Competing interests}
No competing interest is declared.

\section*{Author contributions statement}
MMa, SG and MMo designed the software; MMa developed the first version of the rendering engine. MMa, SG, ND, CPH, VG, VL, JL, YN contributed to the code; ND, JL and GL designed the interface; VG, MMa and MMo wrote the manuscript; MMa, VG, JPP, NL, SM, GL and MMo reviewed the manuscript.

\section*{Acknowledgments}
We thank Prof. SJ Marrink and J Stevens for kindly providing the dataset. 
VTX has received funding from the European Research Council (ERC) under the European Union's Horizon 2020 research and innovation program (grant agreement n° 640283) and from the ERC under the Horizon Europe research and innovation program (grant agreement n° 101069190). CPH and VL are supported by institutional grants
from the National Research Agency under the Investments for the
future program with the reference ANR-18-EURE-0017 TACTIC.
VL is recipient of a fellowship from Qubit Pharmaceuticals. 

%\pagebreak
%Bibliography
\bibliographystyle{unsrt}  
\bibliography{article}

\end{document}